\documentclass[12pt]{article}
\usepackage{amsmath,amssymb}
\usepackage[bookmarks=true]{hyperref}
\usepackage[nosort]{cite}
\usepackage[bulletsep]{collect}
\def\gfxon{\usepackage[final]{graphicx}}

\gfxon

\usepackage{amsfonts}
\usepackage{bm}
\makeatletter
\let\old@startsection=\@startsection
\renewcommand{\@startsection}[6]{\old@startsection{#1}{#2}{#3}{#4}{#5}{#6\mathversion{bold}}}
\makeatother

\def\=:{=\hspace{-.7em}\raisebox{1.1ex}{.}\hspace{.1em}\raisebox{-0.2ex}{.} }

\makeatletter \@addtoreset{equation}{section} \makeatother

\usepackage[width=0.90\textwidth,font={small},labelfont={bf,small}]{caption}

\makeatletter
\def\mr@ignsp#1 {\ifx\:#1\@empty\else #1\expandafter\mr@ignsp\fi}%
\newcommand{\multiref}[1]{\begingroup%\let\protect\string%
\xdef\mr@no@sparg{\expandafter\mr@ignsp#1 \: }%
\def\mr@comma{}%
\@for\mr@refs:=\mr@no@sparg\do{\mr@comma\def\mr@comma{,}\ref{\mr@refs}}%
\endgroup}
\makeatother

\ifx\href\asklfhas\newcommand{\href}[2]{#2}\fi

\begin{document}

\begin{flushright}\footnotesize
%\texttt{ArXiv:\arxivno{1009.xxxx}}\\
\texttt{IFUP-TH/2013-02}, 
\texttt{FTPI-MINN-12/31, UMN-TH-3120/12}
%\vspace{0.5cm}
\end{flushright}
\vspace{0.2cm}
\renewcommand{\thefootnote}{\arabic{footnote}}
	\setcounter{footnote}{0}
\begin{center}%
{\Large\textbf{\mathversion{bold}
Non-Abelian quasigapless Modes\\
Localized on Mass Vortices in Superfluid $^{3}$He-B
}
\par}

\vspace{0.7cm}%

\textsc{Muneto Nitta$^{1}$, Mikhail Shifman$^{2}$ and Walter Vinci$^{3,4,5}$}

\vspace{10mm}
$^1$\textit{Department of Physics, and Research and Education Center 
for Natural Sciences,\\ Keio University, 4-1-1 Hiyoshi, Yokohama, 
Kanagawa 223-8521, Japan}
\vspace{.3cm}
\\
$^2$\textit{William I. Fine Theoretical Physics Institute, University of Minnesota, \\%
 Minneapolis, MN 55455, USA}
\vspace{.3cm}
\\
$^3$\textit{University of Pisa, Department of Physics ``E. Fermi'', INFN,\\%
Largo Bruno Pontecorvo 7, 56127, Italy}\\
$^4$ \textit{London Centre for Nanotechnology, University College London,\\
London, WC1H 0AH, UK.}\\
$^5$ \textit{Department of Computer Science, University College London, \\
London WC1E 6BT, UK.}\\

\vspace{7mm}

\thispagestyle{empty}

\texttt{nitta(at)phys-h.keio.ac.jp}\\
\texttt{shifman(at)physics.umn.edu}\\
\texttt{w.vinci(at)ucl.ac.uk}\\ 

\end{center}

\begin{center}

\textbf{Abstract}\vspace{5mm}

\begin{minipage}{12.7cm}
Kelvin waves, or Kelvons, have been known for a long time as gapless excitations propagating along superfluid vortices. 
These modes can be interpreted as the Nambu-Goldstone excitations arising from 
the spontaneous breaking of the translational symmetry. Recently a different type of gapless excitation
localized on strings  -- the so-called non-Abelian mode -- attracted much attention in high-energy physics. 
We discuss their relevance in condensed matter physics. Non-Abelian rotational quasigapless excitations could
appear  on
the mass vortices in the B phase of the superfluid $^3$He, due to the fact that the order parameter in
$^3$He-B is tensorial. While the $U(1)$ rotational excitations  are well established
in vortices with asymmetric cores, the non-Abelian rotational excitations
belonging to the same family were not considered.

\end{minipage}

\end{center}

\vspace*{\fill}
\section{Introduction}
\label{s1}

In this paper we argue that vortices in certain 
superfluids with a tensorial order parameter, such as the B phase of superfluid $^3$He,
can have non-Abelian quasigapless modes. The existence of such modes depends on the
values of a number of phenomenological parameters [see Eq. (\ref{freen}) below]. Therefore, strictly speaking, 
for the time being we can only say that our assertion refers to a
superfluid  with the
symmetry of $^3$He-B,  characterized by relatively  small values of parameters 
$\gamma_2$ and $\gamma_3$, which could be arranged in ultracold Fermi gases with p-wave pairings.

By non-Abelian we mean that the moduli (collective coordinates) 
that give rise to these modes are described by non-Abelian sigma models, say, the $O(3)$ model. The very notion of non-Abelian gapless (or quasigapless)  modes came from the study of topological defects in high-energy physics, especially in supersymmetric non-Abelian field theories, 
where they attracted much attention recently \cite{Hanany:2003hp,Auzzi:2003fs,Shifman:2004dr,Hanany:2004ea,Eto:2005yh,Eto:2006cx}, 
see also \cite{Eto:2006pg,Shifman:2007ce,Shifman:2009zz} for reviews. 
These non-Abelian modes are called {\it orientational}; they play an important role in the
theory of confinement.

Non-Abelian Nambu-Goldstone (NG) excitations localized on the vortices 
occur when a non-Abelian {\em global} symmetry of the ground state is spontaneously broken by the vortex. 
An interesting case that can be realized in nature is quark matter at high densities,  such as in the neutron star cores. In this case
nucleons melt into the quark matter
in which color $SU(3)$ and flavor $SU(3)$ symmetries are spontaneously broken down to
a diagonal $SU(3)$ color-flavor locked symmetry 
\cite{Alford:2007xm} 
resembling  the B phase of $^3$He.
The vortices in this medium    
spontaneously break  the $SU(3)$ color-flavor locked symmetry of the ground state. 
As a result, non-Abelian NG gapless modes appear, 
localized on and propagating along the non-Abelian vortices  \cite{Balachandran:2005ev,Nakano:2007dr,Nakano:2008dc,Eto:2009kg,Eto:2009bh,Eto:2009tr,Hirono:2010gq,Vinci:2012mc,Cipriani:2012hr}.\footnote{ 
The above modes are gapped, however, after taking into account  quantum corrections \cite{Gorsky:2011hd,Eto:2011mk}.}

Motivated by these developments in high-energy physics we would like to study whether the results obtained there
are relevant in  condensed matter applications.

Superfluid helium-3 is one of the most interesting states of matter thoroughly studied both experimentally \cite{PhysRevLett.28.885,PhysRevLett.29.920} and  theoretically \cite{PhysRevLett.29.1227,RevModPhys.47.331,Leggett:2006,Volovik:2003fe}. Unlike 
conventional superfluids, the fermionic $^{3}$He atoms  condense  after forming bosonic Cooper pairs \cite{PhysRevLett.29.1227}. The attractive interaction between $^{3}$He atoms has a strong short-range repulsive core that favors a $p$-wave pairing and bound states with unit angular momentum and spin. The consequence of this fact is that the order parameter describing the condensate has a tensorial structure: it is a $3 \times 3$ matrix $e_{\mu i}$, where $\mu$  and $i$ are the spin and orbital indices, respectively \cite{RevModPhys.47.331,PhysRevB.36.3583,PhysRevB.24.183}.

Low-energy physics of superfluids can be described by gapless excitations,  the NG modes associated with spontaneously broken global symmetries.  
The most important NG modes are  (i) phonons 
associated with the spontaneously broken phase symmetry $U(1)_{p}$  (here the subscript $p$ stands for phase), and (ii)
magnons associated with spontaneously broken $SO(3)$ spin symmetry  \cite{PhysRev.117.648,Goldstone,PhysRev.127.965,Feynman1958S18}. 

In conventional superfluids the breaking of the Abelian phase symmetry $U(1)_{p}\rightarrow 1 $  leads to the existence of phonon excitations in the bulk and  of topologically stable superfluid vortices \cite{Feynman195517,Abrikosov:1956sx,Onsager} when the superfluid sample is under rotation \cite{RevModPhys.59.533,RevModPhys.59.87,0034-4885-69-12-R03,MineevSalomaa:1986}.   Long ago  Lord Kelvin predicted that vortices  support vibrating modes (called Kelvons) which correspond to  fluctuations of the vortex line \cite{Kelvin,RevModPhys.59.87,PhysRevLett.101.020402}. Kelvons have  been directly observed recently \cite{Fonda:2012}. %arising because of the  breaking of translational and rotational symmetries by the vortex. 
 
In   unconventional superfluids, such as $^{3}$He, the order parameter is  tensorial and non-Abelian symmetries related to spatial rotations are usually  broken in the vacuum.
If we neglect  the spin-orbit interaction, rotations of spinorial and orbital indices can be 
performed independently;  the full symmetry of $^{3}$He is  $G= U(1)_{p}\times SO(3)_{S}\times SO(3)_{L}\,,$
where $SO(3)_{S}$ and $SO(3)_{L}$ are spin and angular rotations. 
Two phases  were theoretically predicted and experimentally (indirectly) observed 
(in the absence of external magnetic fields). We focus on the so-called B phase, where the ground state preserves a locked $SO(3)$ symmetry,
\begin{equation*}
G=U(1)_{p}\times SO(3)_{S}\times SO(3)_{L} \rightarrow H_{\rm B}=SO(3)_{S+L}\,. 
\end{equation*}
 The locked $SO(3)_{S+L}$ symmetry can be identified with the overall
spatial rotations $SO(3)_{R}$. 
The expression above implies the existence of   bulk ``non-Abelian"  NG bosons, called magnons in the condensed matter terminology, due to spontaneous breaking of non-Abelian global symmetries. 
The B phase supports topologically stable vortices:
\begin{eqnarray*}
\pi_{1}\left( U(1)_{p}\times SO(3)_{S}\times SO(3)_{L}/SO(3)_{S+L}  \right)=\pi_{1} \left( U(1)_{p}\times SO(3)_{S-L}  \right)=\mathbb Z \oplus \mathbb Z_{2}\,.
\end{eqnarray*}
The first $\mathbb Z $ factor corresponds to the breaking of the Abelian $U(1)_{p}$ symmetry and supports the so-called mass vortices.
  These vortices are created and stabilized in a  lattice  once the sample is rotated, and are characterized by a nonvanishing superfluid current and angular momentum,  much in the same way  as vortices in  conventional superfluids.\footnote{The second $\mathbb Z_{2}$ factor stabilizes a  more exotic type of vortices, the spin vortices. They are not directly created by the superfluid rotation; they were detected, however,  as bound states with the mass vortices} 
  Both the bulk magnons and the Kelvin excitations  were recently observed \cite{PhysRevLett.90.100403,Hall,PhysRevLett.68.3331,Fonda:2012}  in experiments with superfluid $^{3}$He and $^4$He.
  %\footnote{The Kelvin waves were detected in experiments with the superfluid $^4$He too\cite{Fonda:2012}.} 

In this paper we will argue that a hitherto unexplored type of  quasigapless
modes localized on the mass vortices in the superfluid $^{3}$He-B
exist. While Kelvons can be interpreted as 
the NG modes arising from the breaking of translational symmetry,  the excitations we discuss arise 
from the breaking of the spatial rotation symmetry $H_{\rm B}=SO(3)_{S+L}$ by the vortex solution. Because of their origin we will refer to them as non-Abelian.

It is known that the B phase is divided into two subphases classified according to the core structure of 
the mass vortex under  rotations around the vortex axis. The core   can be either axially symmetric or 
axially asymmetric \cite{PhysRevLett.56.359,PhysRevLett.56.363,PhysRevB.36.3583}.   
A possible signal of
the axial symmetry breaking   in the core of the mass vortices has  been  already observed \cite{Krusius,Pekola,Hakonen}. 
The axial symmetry gives rise to a $U(1)$ NG mode 
localized on the given mass vortex.  This is 
   the conventional $U(1)$ NG excitation, while our task is to focus on non-Abelian excitations.

Considering the whole non-Abelian symmetry,  there can exist up to two or three  additional modes,  due to 
 breaking of the bulk symmetry $H_{\rm B}=SO(3)_{S+L}$ on the vortex, in accordance with the fact that 
\begin{equation*}
 SO(3)_{S+L}/U(1)_z \simeq S^2\quad {\rm or} \quad SO(3)_{S+L}/ 1 \simeq S^{3}/\mathbb Z_{2}
\end{equation*}
 relevant respectively for the axially symmetric core (left) and  for the asymmetric core (right).
 
As far as we know, these  excitations were not   discussed   in the
literature or observed in experiments. This is the first example of spatially localized non-Abelian  modes in condensed matter physics. It is important to stress here that these modes can be quasigapless, or become gapped upon inclusion of quantum effects.

The arguments that lead us to this conclusion are explained in more detail in the next sections. 
They can be applied   to other unconventional superfluids with tensorial order parameters.
The crucial points in our considerations
are (i) the presence  (in the B phase) of an unbroken non-Abelian (global rotational) $SO(3)_{S+L}$ symmetry,
which may be   spontaneously broken by the vortex, and (ii) the tensorial nature of the order parameter.
 The rotational symmetry is broken by the vortex solution at two levels. As in the case of conventional superfluids, it is broken by the physical shape of the vortex itself -- a line defect in three spatial dimensions.
(iii) It is known that the moduli coming from the latter breaking  are not independent of those originated by translations, in the case of scalar order parameter, while we argue that independent (quasi)moduli are generated in the case of tensorial order parameters.

From a more formal point of view, if  the ground state of the theory is symmetric under $H$ and the vortex solution preserves only a subgroup $H'$ of the group $H$, then a particular vortex gives rise to a family of (degenerate) vortices
which are characterized by a number $\nu =  {\rm dim} H - {\rm dim}H'$  of continuous parameters (called moduli) of the solutions and $\nu_{NG}$ NG excitation modes. 
In relativistic field theories, 
the  NG theorem implies the existence of a massless particle per each broken generator ($\nu_{NG}=\nu$).
In the nonrelativistic case \cite{Nielsen:1975hm}   the  NG theorem  is subtler \cite{Watanabe:2012hr,Hidaka:2012ym}:
the number of the gapless excitations can be smaller than or equal to the number of the broken generators (moduli).\footnote{One NG mode with a {\it linear} dispersion relation (called type I) 
corresponds to one broken generator
(as in the relativistic case), and,  at the same time, 
 one NG mode with a {\it quadratic} dispersion relation (called type II) corresponds to {\it two} broken generators. }

Our proposal  is that the rotational symmetry $H=SO(3)_{S+L}$ broken by a nontrivial configuration of the tensorial order parameter on the  vortex solution, generates rotational  non-Abelian (quasi)moduli  $\omega^\ell$ ($\ell =1,2,3$) which are coupled, in the general case, to the  conventional Kelvin excitations. The Kelvin excitation presents an example of the
nonrelativistic type II NG mode.
With two broken translations, there is only one independent circularly polarized fluctuation of the vortex. 
The number of the gapless rotational excitations given a certain number of rotational moduli must be established on the case-by-case 
basis in nonrelativistic theories, through the quantization procedure. 
%
%This paper is organized as follows. 
%In Sec.~\ref{GL} we introduce the Ginzburg-Landau description of the superfluid $^{3}$He 
%and mass vortices in the B phase.
%In Sec.~\ref{conv} we explain why in conventional superfluids, with the scalar order parameter,
%rotational moduli do not arise. 
%For conventional superfluid, the spontaneously broken translational and rotation symmetries of a superfluid vortex produce only translational moduli $\xi_I$ ($I=1,2$ for the $z$ oriented vortex). Furthermore, two broken translational symmetries 
%in the presence of a vortex give rise to one independent Nambu-Goldstone mode, which can be identified with Kelvon. 
%
%At the same time, the spontaneously broken rotational symmetry gives rise to independent  zero modes 
%for the mass vortices provided the order parameter is tensorial, as in superfluid $^3$He-B. In Secs.
%\ref{mv} and \ref{mvwem}
%we show that there exist two or three gapless modes with linear dispersion   
% localized on the mass vortex 
%with symmetric and asymmetric cores, respectively.  
%In Sec.~\ref{sec:LEET} we outline the construction of  the low-energy effective theory of the gapless excitations 
%localized on the vortex. Such effective theories are defined as 1+1 dimensional theories. 
%Sec.~\ref{sec:conclusion} summarizes our conclusions. 
%

\section{The Ginzburg-Landau description of $^{3}$He}
\label{GL}

The starting point of our analysis is the Ginzburg-Landau description of the superfluid phase of  $^{3}$He near the critical temperature. The order parameter is a $3\times 3$ matrix $e_{\mu i}$ with spin indices $\mu$ (where $\mu=1,2,3)$ and  orbital indices $i \,\,\, (i=1,2,3)$ that transform under the vector representations of $SO(3)_{S}$ and $SO(3)_{L}$,
\begin{equation}
 e_{\mu i}\rightarrow e^{i \psi} S_{\mu \nu} L_{ i j} e_{\nu j}\,,
\end{equation}
where $S$ and $L$ are $SO(3)_{S}$ and $SO(3)_{L}$ orthogonal rotation matrices. $\psi$ is a phase parameter corresponding 
to $U(1)_{p}$ rotations. 
  
The most general  time-dependent  expression for  the free energy, invariant under the $U(1)\times SO(3)_{S}\times SO(3)_{L}$ symmetry of the theory takes the form
\begin{align}
&F_{GL} \,  \equiv  \,\,\hbar i\, e_{\mu i}\partial_{t}e^{*}_{\mu i}+ \gamma_{1}\partial_{i}e_{\mu j}\partial_{i}e^{*}_{\mu j}+\gamma_{2}\partial_{i}e_{\mu i}\partial_{j}e^{*}_{\mu j}+\gamma_{3}\partial_{i}e_{\mu j}\partial_{j}e^{*}_{\mu i}-\alpha \, e^{*}_{\mu i}e_{\mu i}&\,\nonumber \\[3mm]
&+ \beta_{1}  e^{*}_{\mu i}e^{*}_{\mu i} e_{\nu j}e_{\nu j}+ \beta_{2}  e^{*}_{\mu i}e_{\mu i} e^{*}_{\nu j}e_{\nu j} 
+  \beta_{3}  e^{*}_{\mu i}e^{*}_{\nu i} e_{\mu j}e_{\nu j}+\beta_{4}  e^{*}_{\mu i}e_{\nu i} e^{*}_{\nu j}e_{\mu j}+ \beta_{5}  e^{*}_{\mu i}e_{\nu i} e_{\nu j}e^{*}_{\mu j} \,.&
 \label{freen}
\end{align}
The coefficients $\gamma$ and $\beta$ are phenomenological parameters depending both on temperature and pressure. Experimental determination of all these coefficients is very difficult \cite{PhysRevB.75.174503}. One can roughly evaluate them from BCS-like calculations \cite{PhysRevB.24.183}. Note that the $\gamma_2$ and $\gamma_3$ terms 
in the second line differ only by a total derivative. Hence, in deriving the equations of motion 
in the bulk they can be identified.
 BCS-like calculations are valid at weak coupling, and predict the B phase, were they are more reliable. Strong coupling effects have to be taken into account in general, especially to predict the appearance of the A phase. However, for the considerations presented in this paper the precise values of the coefficients $\gamma ,\,\alpha$ and $\beta$  are not important.

Minimizing the potential term in Eq.~(\ref{freen}) with respect to the order parameter gives the B-phase ground state  \cite{Volovik:2003fe} 
\begin{eqnarray}
\left(e_0\right)_{\mu i }=e^{i\psi}\Delta \left(R_0\right)_{\mu i}\,,\quad \quad  \Delta^{2}\equiv\frac{\alpha}{6(\beta_{1}+\beta_{2})+2(\beta_{3}+\beta_{4}+\beta_{5})}\,,
\label{eq:vacuuma}
\end{eqnarray}
where the ground state solution is
parametrized (in the most general form)  in terms of an element $\left(R_0\right)_{\mu i}$ of the  $SO(3)$ group.
The configuration$\left(e_0\right)_{\mu i }=e^{i\psi}\Delta \left(R_0\right)_{\mu i}$  contains four parameters, representing the NG excitations in the bulk, or magnons, discussed in the Introduction.

The unbroken symmetry corresponds to a combined action of spinorial and orbital transformations, namely,
\begin{equation}
 S_{ \mu \tilde\mu} L_{i\tilde i}\left(R_0\right)_{\tilde\mu \tilde i} = \left(R_0\right)_{\mu i}\, .
 \label{27}
\end{equation}
Without loss of generality we can set $\left(R_0\right)_{\mu i}=\delta_{\mu i}$ (i.e. $\theta =0$ and $\psi=0$) in the ground state. 
Then Eq. (\ref{27}) becomes obvious provided $S=L$ .

If the parameters $\gamma_{2,3}$ in (\ref{freen}) are small, the symmetry of the free energy (\ref{freen}) 
is enhanced.
Indeed, in  the limit $\gamma_2,\gamma_3 \to 0$, the spatial $SO(3)_{R}$  rotations of coordinates can be done independently from  
the spinorial $SO(3)_S$ and orbital rotation $SO(3)_L$ without changing the free energy. In this case, the $SO(3)_S$ and $SO(3)_L$ symmetries, together with the locked $SO(3)_{S+L}$ symmetry effectively act as internal symmetries.
Then if we switch on $\gamma_{2,3}  \neq 0$ these two terms in the free energy explicitly break
 $SO(3)_{\rm S}\times SO(3)_L\times SO(3)_R$ down to a  
 ``locked"  $SO(3)_{S+L}=SO(3)_{R}$. 
This observation  will be important below.

\subsubsection*{Vortices in $^3$He-B}
\label{mv}

The simplest topologically stable mass vortex in the B phase of $^{3}$He has the  following asymptotic behavior
 \begin{equation}
e^{\rm vort}_{\mu i}=\Delta \,e^{i \alpha} \delta_{\mu i}\,\,\,\mbox{at}\,\,\, r\to\infty\,,
\label{bound}
\end{equation}
where $\alpha$ is the polar angle in the perpendicular plane. We identify $\psi$ with the polar angle, so
that the  phase of the order parameter
 winds by $2\pi$. 
From the topological point of view, a mass vortex is stable because $\pi_1(U(1)_p)$ is nontrivial.

The asymptotic behavior (\ref{bound}) of the vortex solution does not break the $SO(3)_{L+S}$ symmetry of the bulk. 
Thus, no modulus associated with the phase winding appears.
At generic distances, the ansatz (\ref{bound}) is generalized as follows:
\begin{equation}
e^{\rm vort}_{\mu i}(x,y)=\Delta \, e^{i\alpha} \,\tilde e_{\mu i}(x,y)\,, \quad
\tilde e_{\mu i}\equiv
\left(
\begin{array}{ccc}
 g_{1}(x,y) & f_{12}(x,y)  & f_{13}(x,y)  \\
f_{21}(x,y)  & g_{2}(x,y)  & f_{23}(x,y)  \\
 f_{31}(x,y) & f_{32}(x,y)  & g_{3}(x,y)  
\end{array}
\right),
\label{mat}
\end{equation}
with boundary conditions 
$\tilde e_{\mu i}\to \delta_{\mu i}\, \,\,\,\mbox{at}\,\,\, r\to\infty\,.$

 As already mentioned in the Introduction, vortices with an asymmetric core and a rotational modulus can exist. They are characterized by a profile matrix $\tilde e_{\mu i}(x,y)$ that breaks  $SO(2)$ rotations around the vortex axis,  for example if all $f$'s vanish, but $g_1\neq g_2$ in the core. This is a well-known property that distinguishes the  mass vortices in $^{3}$He-B  from those in conventional superfluids \cite{PhysRevLett.56.359,PhysRevLett.56.363}. Thus,  the $^{3}$He vortices cannot be considered as diagonal embeddings of more conventional $U(1)$ superfluid vortices. 

The mentioned rotational gapless mode is Abelian. In the following, we want to consider its non-Abelian counterpart.

\section{Scalar vs. nonscalar order parameters}
\label{conv}

To begin with, let us consider the usual case of a superfluid vortex in a theory with scalar order parameter
(e.g. in $^4$He). Such vortices are characterized by two translational moduli, corresponding to translations
in the $x$ and $y$ directions (assuming that the vortex axis is oriented along the $z$ axis). Let us examine the physical reason behind the existence of these translational moduli (and, hence, the Kelvin mode).
The underlying theory  has a number of invariances in the ground state. In the case of the scalar order parameter
these symmetries are (a) translational (three generators) and  (b) rotational (three generators). Some of the above symmetries are obviously spontaneously broken by the vortex defect, namely, translations in the $x$ and $y$ direction ($T_{x,y}$) and rotations outside the $x,y$ plane ($L_{zx}$ and $L_{zy}$).  

It turns out that the two translational modes are one the conjugate momentum of the other, thus giving a single mode known as the Kelvin mode (see e.g. \cite{Nielsen:1975hm}).  
What is the fate  of the rotational symmetry which is also spontaneously broken by the vortex? Two of
the rotations are broken, while the $SO(2)$ rotation around the $z$ axis remains unbroken. At first sight the $SO(3)/SO(2)$
pattern of the symmetry breaking must lead to two extra moduli, corresponding to broken $L_{zx}$ and $L_{zy}$.
However, these extra moduli are absent because the above rotations are already represented by
the translational 
moduli,\footnote{In the context of high-energy physics this observation is sometimes referred to as 
the inverse Higgs mechanism \cite{Low:2001bw,Ivanov:1975zq,Clark:2002bh}} 
which we will call $\xi_I$ (where the capital latin subscript runs over $I=1,2$). In the quantization procedure the moduli $\xi_I$ become the fields $\xi_I (t,z)$. Thus, in the case at hand, no extra moduli are needed to represent
the spontaneously broken $O(3)$ symmetry since local translations are equivalent to local rotations \cite{Low:2001bw}.

Dynamics of the $\xi_I$ on the vortex
is described by the gradient terms $\sim(\partial_z \xi_I)^2$, plus a time derivative. 
The string oriented along the $z$ axis corresponds to the  ground state:
\begin{eqnarray}
\xi_I =0\,.
\label{anvap}
\end{eqnarray}
A rotated string is described by another ground state,
\begin{eqnarray}
\xi_I (z) = a_I\, z \,,\quad \mbox{ $a_I$ arbitrary real numbers}.
\label{anva}
\end{eqnarray}
This is also a solution of the equation of motion. An apparent increase in the ground state energy for 
(\ref{anva}) compared to (\ref{anvap}) is an illusion: for the rotated vortex the element of distance along the vortex changes too. 
The overall tension remains the same.

Let us pass to the problem of interest in which the relevant order parameter carries spatial indices, see Eq. (\ref{freen}). 
We will  start our discussion  from the simplest limit $\gamma_{2,3}=0$  (or very small).
If $\gamma_{2,3}=0$ the symmetry of the ground state is enhanced:
in addition to the translational symmetry and that of spatial rotations $R$, it has ``internal" symmetries $SO(3)_S\times SO(3)_L$   spontaneously broken in the ground state
down to $SO(3)_{L+S}$. 

The vortex solution itself breaks $T$ and $R$ spontaneously, by the orientation of the vortex axis.
As was explained above, this breaking generates only two (not four) moduli, $\xi_I$. 
The breaking of the vacuum $SO(3)_{L+S}$ symmetry (which in the above limit is ``internal")
necessarily produces either three rotational moduli in the vortex solution [if $SO(3)_{L+S}$ is completely broken on the given solution]
or two such moduli [if $SO(3)_{L+S}$ is broken down to $SO(2)$].
For definiteness let us focus on the first case.
If the axially asymmetric solution exists in the  limit $\gamma_{2,3}=0$, it can be automatically elevated to the status of the full-fledged non-Abelian vortex. Indeed, since it breaks $SO(2)$ by definition, one can ``smear" this $SO(2)$ over the $SO(3)_{L+S}$ by considering all possible embeddings of the broken $SO(2)$ in $SO(3)_{L+S}$. In this way, we acquire three (non-Abelian) rotational moduli. The axially asymmetric solution exists provided that either $g_1\neq g_2$, or one or more $f$'s in Eq. (\ref{mat}) do not vanish, or both. 

We did not attempt to systematically explore the parameter space $\beta_{1, ..., 5}$ in  search of all 
possible global vortex solutions 
at $\gamma_{2,3}=0$. This will be the subject of a special investigation.
However, for some values of $\beta_{1, ..., 5}$ particular relevant solutions can be found in a qualitative way
in the following ansatz: 
\begin{equation}
e^{\rm vort}_{\mu i}=\Delta (r_\perp) \,e^{i \alpha} \delta_{\mu i} + \varepsilon_{\mu ik}\,\chi^k (r_\perp)
\end{equation}
where $\vec r_\perp =\{x, y\}$ and $\chi^i$ is a vector with regards to the unbroken $SO(3)_{L+S}$. In addition to
(\ref{bound}), $\chi^k (r_\perp)\to 0$ at $r_\perp\to\infty$. In the core of the vortex, at small $r_\perp$,
an instability for $\chi^i$ develops, $\chi^i\neq 0$, and we obtain two rotational moduli [corresponding to
$SO(3)\to SO(2)]$.

Now let us consider a  general situation and switch on $\gamma_{2,3}\neq 0\,.$
Applying $SO(3)_{L+S}=SO(3)_R$ to the vortex profile  we get, generally speaking,  three moduli $\omega^\ell$ in the matrix structure of 
$e^{\rm vort}_{\mu i}(x,y)$. Simultaneously the center of the solution gets shifted too by $\xi_I$, see Eq. (\ref{anva}), due to the entanglement of  rotations and translations. The shift parameters $a_I$ are linearly dependent on $\omega_{1,2}$. 

There are two consequences: a mass gap proportional to $\gamma_{2,3}$ is generated
for those moduli that are not associated with the $U(1)$ rotation and, in addition, a derivative coupling connecting the translational and rotational moduli appears too.

\section{Low-energy effective theories on the vortex }
\label{sec:LEET}

In this section we will briefly outline a general construction of low-energy effective theories 
on the vortex which is not very widely used in condensed matter (albeit is absolutely standard in
high-energy physics, e.g. \cite{Manton:1981mp}). This theory allows one to describe dynamics of gapless 
and quasigapless excitations. This technique will help us to substantiate the last statement in Sect. \ref{conv}.

The general prescription for deriving an effective low-energy theory  is as follows. One  expands the Ginzburg-Landau free energy (\ref{freen}) in terms of  fluctuations of the order parameter, up to the second order in derivatives, near the vortex solution. These fluctuations are assumed to be dependent on the world-sheet coordinates of the vortex,
\begin{equation}
e_{\mu i}(x,y) = e^{\rm vort}_{\mu i}(x,y)+\delta e_{\mu i}(x,y,z,t)\,.
\end{equation}
The variation above is still completely generic.
When varying expression (\ref{freen}), we observe that the terms proportional to the first order in the variations vanish because of the equations of motion. The remaining terms  are as follows:
\begin{align}
\delta^{2} \left( F_{GL}
\right)& =  \hbar i\, \delta e^{*}_{\mu i}\partial_{t}\delta e_{\mu i}+ \gamma_{1}\partial_{z}\delta e_{\mu j}\partial_{z}\delta e^{*}_{\mu j}+\gamma_{2}\partial_{z}\delta e_{\mu z}\partial_{z}\delta e^{*}_{\mu z}+\gamma_{3}\partial_{z}\delta e_{\mu z}\partial_{z}\delta e^{*}_{\mu z}+\nonumber \\[2mm]
& +  \gamma_{2}\partial_{z}\delta e_{\mu z}\partial_{I}\delta e^{*}_{\mu I}+\gamma_{z}\partial_{z}\delta e_{\mu I}\partial_{I}\delta e^{*}_{\mu z}+{\rm c.c.}+\nonumber \\[2mm]
 & +  \gamma_{1}\partial_{I}\delta e_{\mu J}\partial_{I}\delta e^{*}_{\mu J}+\gamma_{2}\partial_{I}\delta e_{\mu I}\partial_{J}\delta e^{*}_{\mu J}+\gamma_{3}\partial_{I}\delta e_{\mu J}\partial_{J}\delta e^{*}_{\mu I}+(\partial_{e^{*}_{\mu i}} \partial_{e_{\nu j}} V )\delta e_{\mu i} \delta e^{*}_{\nu j} + ...\,,
 \label{eq:fulllag}
\end{align}
where the ellipses denote terms of the third and higher order in $\delta e_{\mu i}(x,y,z,t)$  containing at most two derivatives.
We will discuss them later.

For the time being we discard the $t$ and $z$ dependences in (\ref{eq:fulllag}) and vary $\delta^{2} \left( F_{GL}\right)$ with respect to $\delta e_{\mu i}$.
In this way we arrive at a static differential operator  acting on $\delta e_{\mu i}$, and define its eigenvalues and eigenmodes,
\begin{eqnarray}
L(x,y) e_{\mu i}^{(n)} (x,y) = E^{(n)} e_{\mu i}^{(n)} (x,y)\,.
\end{eqnarray}
Then one uses adiabatic approximation to expand $\delta e_{\mu i}(x,y,z,t) =\sum_n c_n(t,z) e_{\mu i}^{(n)} (x,y)$.
Of all infinite set of modes we are interested only in the gapless and quasigapless modes, i.e. those for which $E^{(n)}=0$ or small. Again, it is convenient to start from the limit $\gamma_{2,3} =0.$

Let us denote the set of moduli on which $e^{\rm vort}_{\mu i}(x,y)$ depends as $m^a$. This set includes
two translational moduli $\xi _I$ and three orientational $\omega^\ell$. Then
 $e^{\rm vort}_{\mu i}(x,y; m^a)$
determines the gapless modes,
\begin{eqnarray}
e_{\mu i}^{(0,a)} (x,y)\propto \frac{\partial}{\partial {m^a}} \,e^{\rm vort}_{\mu i}(x,y; m)\,,\qquad E^a =0\,,
\label{66}
\end{eqnarray}
where the proportionality coefficient is derived from the normalization of each mode.
Substituting these gapless modes in the expansion of $\delta e_{\mu i}(x,y,z,t)$ (and neglecting all nongapless modes),
\begin{eqnarray}
\delta e_{\mu i}(x,y,z,t) = \sum_a m^a(z,t) e_{\mu i}^{(0,a)}\,,
\end{eqnarray}
we arrive at a low-energy effective theory on the vortex in the form
\begin{eqnarray}
F_{\rm LE}^{\rm grad} = \sum_{a,b} G_{a,b} (m) \partial_z m^a(z,t) \partial_z m^b(z,t)
\end{eqnarray}	
plus a similar term with time derivative (quadratic in $\partial_t$ for the type I NG modes and linear for the type II NG modes). In the quadratic approximation, as in (\ref{eq:fulllag}), the target space metric
$G_{a,b} (m)$ is replaced by $\left.G_{a,b} (m)\right|_{m^a\equiv 0}$. The overall $m$ dependence of $G_{a,b} (m)$ is fixed by the pattern of the symmetry breaking, or, alternatively, algebraically
\begin{eqnarray}
G_{a,b} (m) = \int d^2x_\perp  \frac{\partial e^{\rm vort}_{\nu j}(x,y; m)}{\partial m^a} \,
\frac{\partial e^{\rm vort}_{\mu i}(x,y; m)}{\partial m^b}
\label{67}
\end{eqnarray}
where convolutions of the spin and angular momentum indices are ignored, so that Eq. (\ref{67}) is somewhat symbolic.
Usually, translational moduli are uncorrelated with others.
Then the metric (\ref{67}) factors out and acquires a block form 
%(see e.g. [LSV], this is the case when $\gamma_{2,3} = 0$)
\begin{eqnarray}
F_{\rm LE} = \frac{T}{2} \partial_z \xi_I \partial_z \xi_I +
\sum_{\ell,\tilde\ell} G_{\ell,\tilde\ell} (\omega) \partial_z \omega^\ell (z,t) \partial_z \omega^{\tilde\ell}(z,t) +
\mbox{time der.}
\label{68}
\end{eqnarray}	
In the previous sections we explained that our present problem is peculiar since the family of solutions is obtained under simultaneous and concerted
variations of rotational and translational moduli. 
	
Now, let us switch on $\gamma_{2,3} \neq 0$ but small. Since $SO(3)$ in this case is explicitly broken
by the vortex axis, only the modulus corresponding to the axial $SO(2)$ will remain strictly
gapless. Two others -- belonging to the coset $SO(3)/SO(2) $ -- generally speaking will acquire a mass gap, proportional to $\gamma_{2,3}$, corresponding to a small shift of two eigenvalues in (\ref{66}). In other words, we expect
$\Delta F_{\rm LE,m} \propto \gamma_{2,3} \sum_{\ell_\perp} \mu_{\ell_\perp\tilde{\ell_\perp}} 
(\omega^{\ell_\perp}\omega^{\tilde\ell_\perp})$ ($\ell , \tilde \ell = x,y$).
We will show elsewhere that these modes are type-I NG modes with 
a linear dispersion relation so that all modes are independent.
Inclusion of quantum effects (not discussed here) 
may or may not lift the $U(1)$ mode too, depending
on dynamical details.

In addition to
(\ref{68}) we will have an extra term $\Delta F_{\rm LE}^{\rm grad} \propto \gamma_{2,3} \, (\partial_z \xi_I)\,( \partial_z \omega^I)$. 
There is a dissipation of energy from the Kelvin mode to rotational and vice versa with the coefficient
$\sim\gamma_{2,3}$.

\section{Conclusions} 
\label{sec:conclusion}

In this paper, we discussed  the emergence of non-Abelian (quasi)gapless modes on the vortices in $^3$He.
So far our analysis is purely classical. Quantization needed to determine how many (quasi)gapless excitations 
appear from given moduli, and whether or not some are lifted by quantum effects is deferred
till a more detailed publication.\footnote{In the case of non-Abelian vortices in dense QCD,
gapless modes localized on the vortex are 
gapped after taking into account quantum corrections; 
quantum magnetic monopoles show up \cite{Gorsky:2011hd,Eto:2011mk}.}

In conventional superfluids (e.g. $^4$He) the spontaneously broken translational and rotational symmetries 
on a  vortex do not produce independent NG modes; one can say
that only translational excitations occur, 
a single  NG excitation, 
the Kelvon. This is due to the scalar nature of the order parameter.

At the same time, 
for the mass vortices in the superfluid $^3$He-B, 
the rotational symmetry gives rise to independent  rotational (quasi)NG excitations.
One of them -- due to the axial asymmetry of the vortex core -- was known for a long time.
We argued that extra non-Abelian rotational excitations exist,
associated with the full  $SO(3)$ 
symmetry of the ground state. 
Generically, they are not strictly gapless and can be derivatively coupled with the translational excitations, the Kelvons.
Whether or not they will show up as quasigapless depends on 
particular values of $\gamma_{2,3}$ and $\beta_{1, ... , 5}$. Generically,  the parameters $\gamma_2$ and
$\gamma_3$ are not parametrically small compared to $\gamma_1$ in $^3$He-B , and the gap of the new modes could be in principle of the order of the fermionic gap. As already mentioned in the Introduction, this problem can be in principle cured in ultracold Fermi gases with p-wave pairing.

The tensorial order parameter in $^3$He-B is crucial for the phenomenon.
We outlined the  construction of the low-energy effective theory for the (quasi)gapless excitations 
localized on the vortex,  defined as a (1+1)-dimensional theory on the vortex world sheet.

\section*{Acknowledgements}
We thank 
Takeshi Mizushima, Yasumasa Tsutsumi for comments
and Gregory.~E.~Volovik for discussions.
This work of M.N. is supported in part by 
Grant-in-Aid for Scientific Research (No. 23740198) 
and by the ``Topological Quantum Phenomena'' 
Grant-in-Aid for Scientific Research 
on Innovative Areas (No. 23103515)  
from the Ministry of Education, Culture, Sports, Science and Technology 
(MEXT) of Japan. The work of M.S. is supported in part by DOE grant DE-FG02- 94ER-40823.

\newpage

 \bibliography{BibHelium}
\bibliographystyle{nb}

\end{document}